\documentclass[11pt]{article}

\usepackage{amssym}
\usepackage{psfig}
\usepackage{times}
\usepackage{wrapfig}
\makeatletter
\renewcommand{\section}{\@startsection{section}{1}{0in}
	{0.4\baselineskip}{0.1\baselineskip}{\Large\bf}}
\renewcommand{\subsection}{\@startsection{subsection}{2}{0in}
	{0.25\baselineskip}{-\baselineskip}{\large\bf}}
\renewcommand{\subsubsection}{\@startsection{subsubsection}{3}{0in}
	{0.1\baselineskip}{-\baselineskip}{\normalsize\bf}}
\makeatother
\newcommand{\jpg}{J. Phys. G.: Nucl. Part. Phys.}

\newcommand{\fullcircle}\bullet
\newcommand{\opencircle}\circ
\newcommand{\opentriangledown}\bigtriangledown

\begin{document}

%
\thispagestyle{myheadings}
%
\markright{OG 4.3.11}
\begin{center}
%
{\LARGE \bf Geomagnetic Effects on the Performance of Atmospheric \v{C}erenkov Telescopes}
\end{center}

\begin{center}
%
%
{\bf  P.M. Chadwick, K. Lyons, T.J.L. McComb, K.J. Orford, J.L. Osborne, S.M.
Rayner, S.E. Shaw, and K.E. Turver\\}
{\it Dept. of Physics, Rochester Building, Science
Laboratories, University of Durham, Durham DH1 3LE, UK}
\end{center}

\begin{center}
{\large \bf Abstract\\}
\end{center}
\vspace{-0.5ex}
%
%
Atmospheric \v{C}erenkov telescopes are used to detect electromagnetic
showers from primary gamma rays of energy $ > 300$ GeV and to
discriminate these from cascades due to hadrons using the shape and
orientation of the \v{C}erenkov images. The geomagnetic field affects
the development of showers and diffuses and distorts the images. When
the component of the field normal to the shower axis is sufficiently
large ($> 0.4$ G) the performance of gamma ray telescopes may be
affected. 
%

\vspace{1ex}

%
%


\section{Introduction}

Bowden et al. (1991) discussed the effect of the geomagnetic field on
the performance of ground based gamma ray telescopes. The interaction of
the field and the cascade electrons produces a broadening of the
atmospheric \v{C}erenkov light image resulting in a reduction in the
density of light sampled by the telescope; so the energy threshold for
the telescope increases. This is observed in the higher counting rate
for a telescope when detecting showers propagating along the lines of
the field (with no spreading) than when observing cascades developing
perpendicular to the field lines (and being spread). Typical differences
in measured count rate were about 20\% in these extreme cases. The
possibility was noted, on the basis of simulations, that an associated
rotation of the direction of the \v{C}erenkov light images may occur for
cascades developing under high magnetic fields and in unfavourable
directions. This could be of importance in ground-based gamma ray
studies since the orientation of the image in gamma ray initiated
cascades is the key to rejection of $>$ 99\% of the charged cosmic ray
background (Hillas, 1985).
 
Lang et al. (1993) showed that their measurements of TeV gamma rays from
the Crab nebula using the imaging \v{C}erenkov technique were not
significantly affected by magnetic fields $<$ 0.35 G.

We report measurements made using a ground based gamma ray telescope of
cascades developing under the influence of fields up to 0.55 G.
Observations with the Mark 6 telescope operating in Narrabri, Australia
which is discussed by Armstrong et al. (1999) are subject to such
magnetic fields when observing objects to the south. The observational
data demonstrate all the effects of the geomagnetic field on cascades
which are predicted by simulations.


\section{Experimental results}

Most of the data considered here were taken during routine observations
of potential gamma ray sources during 1996 -- 1998. A small amount of
data was taken in dedicated measurements at fixed zenith angles of
$40^\circ$ and a range of azimuth, corresponding to a range of values of
the geomagnetic field. The only restriction imposed on data was that
events should lie within $1^\circ$ of the centre of the camera (to avoid
edge effects) and should be large enough (5 times the triggering threshold)
to ensure that their shapes were well measured.

\begin{wrapfigure}[21]{l}{10.2cm}
\centerline{\psfig{file=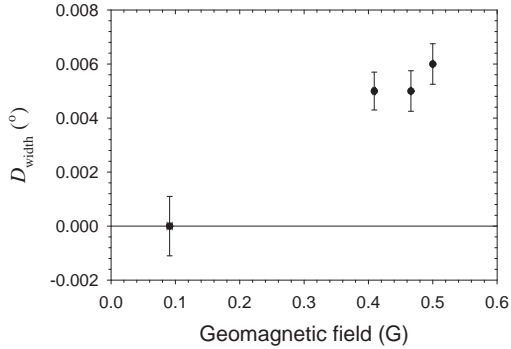,height=6cm}}
\caption{The observed difference in the mean width ($D_{\rm width}$) for
images in hadronic cacades developing parallel and perpendicular to
fields of varying strength for zenith angles in the range $35^\circ -
40^\circ$. The concentration of data at values of high magnetic field
strength reflects the directions of observation of potential VHE sources
from Narrabri.} \label{fig:distortion}
\end{wrapfigure}

The widths of images recorded by the Mark 6 telescope in directions
parallel and perpendicular to the magnetic field have been investigated.
For a sample of cascades measured during observations of a range of
sources the small but significant differences of the mean width in the
parallel and perpendicular directions are shown in figure
\ref{fig:distortion} for a range of values of distorting field. The
widths of cascades in directions perpendicular to the field are
significantly greater than those in parallel directions for fields $>
0.4$ G. Although the differences in widths are small because of the
effects of pixellation and noise which are common to all data, the
measurements are free from systematic effects. This is because the
orientation of images to the magnetic field depends on the orientation
of the image in the camera. Measurements of the width of the images
parallel and perpendicular to the magnetic field are derived from events
distributed throughout the same observation.

In the absence of any magnetic field and detector triggering biases, the
distribution of the orientation of the images in the camera of a
telescope will be isotropic. If we define a coordinate system where the
angle $a_{\rm x}$ is the angle of the long axis of an image to the
horizontal in the camera frame, then the distribution of $a_{\rm x}$
should be flat between $-90^{\circ}$ and $90 ^{\circ}$. Minor deviations
from uniformity in angle might be expected near to threshold because of
the increased effect of small changes in triggering probability and the
six-fold symmetry of the hexagonal arrangement of close-packed
detectors.

\begin{figure}[b!]
\centerline{\psfig{file=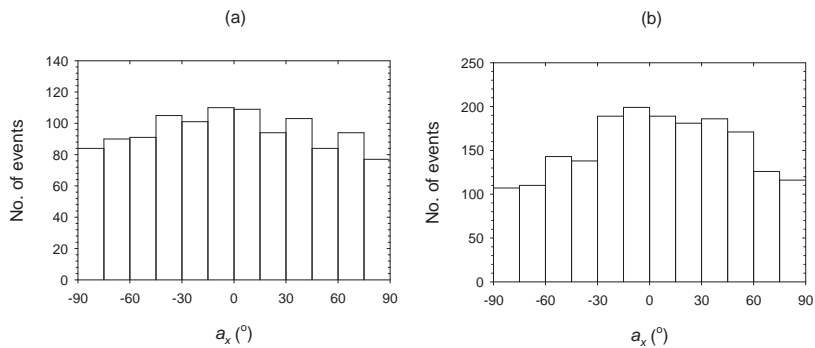,height=6cm}}
\caption{Distribution of $a_{\rm x}$ for cascades developing under (a) a
small transverse geomagnetic field ($B = 0.15$ G) and (b) a
large transverse geomagnetic field ($B = 0.52$ G). \label{fig:aysmallh}}
\end{figure}

We show in figure~\ref{fig:aysmallh}(a) the distribution in the angle
$a_{\rm x}$ for background hadronic events which are subject to small
values of the transverse magnetic field (0.15 G). We show in
figure~\ref{fig:aysmallh}(b) the distribution in $a_{\rm x}$ for images
due to hadron-induced cascades recorded under a transverse geomagnetic
field of 0.52 G. All of these events were recorded in dedicated
observations with the telescope at a fixed zenith angle ($40^\circ$).
Data at different values of the transverse magnetic field were obtained
by varying the azimuth angle.
 
All data in figure~\ref{fig:aysmallh} were recorded within a period of
one hour, so potential variations due to camera performance, atmospheric
clarity, etc. were minimised. The data were processed following our
normal procedures (see e.g. Chadwick et al., 1998b). The requirement
that images have a minimum brightness and fall within the camera ensures
that the data are free of effects of variations in night sky brightness.
The data for $B = 0.15$ G show the expected distribution indicative of a
near isotropic distribution of directions in the camera with a small
peak. We note a strong anisotropy, with a peak containing twice the
minimum number of events resulting from the skewing or distortion of the
image, for events subject to a $0.5$ G field. The maximum of the
distortion occurs at $a_{\rm x} = 0$, which is appropriate for these
observations which were made at magnetic south.

\begin{wrapfigure}[17]{l}{8cm}
\centerline{\psfig{file=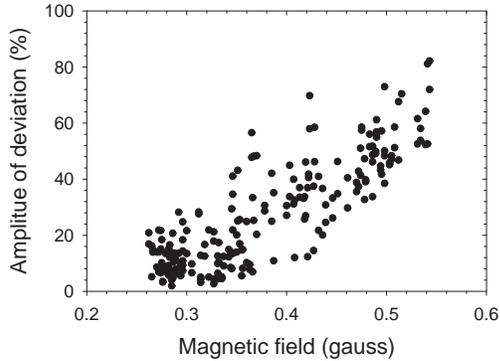,height=6cm}}
\caption{The measured amplitude of $a_{\rm x}$ anisotropy as a function of
transverse geomagnetic field.} \label{fig:ayampvh}
\end{wrapfigure}
The magnitude of the anisotropy displayed in
figure~\ref{fig:aysmallh}(b) should depend on the strength of the
magnetic field. In figure~\ref{fig:ayampvh} we plot the amplitude of the
peak of the distortion in the $a_{\rm x}$ distribution as a function of
the magnetic field strength. Each point corresponds to the result for a
15 minute segment of data taken as part of routine telescope operation.

Note that for values of transverse component of the geomagnetic field
less than 0.35 G there is no great distortion of the $a_{\rm x}$
distribution, as suggested by the work of Lang et al. (1993), but for
values of field in excess of 0.4 G substantial distortion occurs.

\begin{wrapfigure}[19]{l}{8cm}
\centerline{\psfig{file=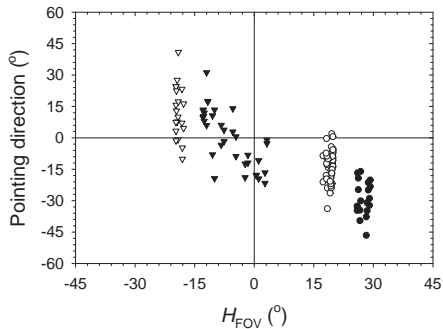,height=5.5cm}}
\caption{The measured direction of $a_{\rm x}$ anisotropy as a function
of the direction of transverse geomagnetic field. $\opentriangledown$
represents data from the directions of Cen X-3, $\blacktriangledown$
from SMC X-1, $\opencircle$ from PKS 2005--489 and $\fullcircle$ from
PKS 2155--304 to cover a range of $H_{\rm FOV}$.} \label{fig:aydirvh}
\end{wrapfigure}

The position of the peak in the $a_{\rm x}$ distribution depends on the
angle between the projected magnetic field and the vertical direction in
the camera --- $H_{\rm FOV}$. We show in figure~\ref{fig:aydirvh} the
correlation between $H_{\rm FOV}$ and the position of the peak in
$a_{\rm x}$. The data demonstrate the expected relation between these
angles.


\section{The orientation of gamma ray images}

A number of potential gamma ray sources has been observed using the Mark
6 telescope at Narrabri. In several cases there is evidence for gamma
ray emission (Chadwick et al., 1998a, 1998b, 1999). The evidence comes
mainly from a comparison of the {\it ALPHA} distributions for data
selected on the basis of image shape for the ON-source and OFF-source
scans. The difference between the {\it ALPHA} distributions should show
an excess of events --- the gamma ray candidates --- at low values of
{\it ALPHA}. If the data were taken in geomagnetically unfavourable
directions, as is the case for most of our data, it might be expected
that the {\it ALPHA} distribution of the excess events would be wider
than that for data taken in more favourable geomagnetic directions.
Consideration of data recorded from a gamma ray source which are subject
to $B < 0.35$ G may provide the only true indication of the {\it
ALPHA}-distribution for gamma rays in our telescope.

\begin{figure}[t!]
\centerline{\psfig{file=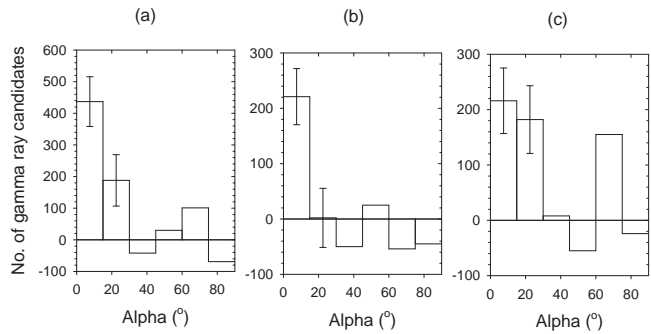,height=5.0cm}}
\caption{The distribution in {\it ALPHA} of excess gamma ray events from
PKS 2155--304. (a) is for all events, (b) those for transverse fields
$<$ 0.35 G. and (c) those for transverse fields $>$ 0.35 G.}\label{fig:alphaplot2}
\end{figure}

Observations of PKS 2155--304 were made with the cascades recorded over a
range of transverse geomagnetic field strengths between 0.25 and 0.5 G.
The total data set contained 41 hrs of observation, as reported
(Chadwick et al. 1999). The {\it ALPHA} plot for the difference between
the ON-source and OFF-source data taken at zenith angles $\theta <
45^{\circ}$ is shown in figure~\ref{fig:alphaplot2}(a). The significance
of the excess at {\it ALPHA} $< 30^\circ$ is $5.7 \sigma$. We are able
to select a subset of data for which the strength of the projected field
to which the cascades were exposed was $<$ 0.35 G and for which minimal
distorting effect would be expected. The {\it ALPHA} plot for this
subset is shown in figure~\ref{fig:alphaplot2}(b). The excess events all
have {\it ALPHA} $ < 15^\circ$ and the distribution is narrower than
that of the total dataset and is typical of that expected for a $0.25
^{\circ}$ pixel camera. (In the absence of any other data for gamma rays
detected with our telescope and {\it not} subject to the effects of the
magnetic field, we assume that this is reasonable.)

The {\it ALPHA} plot for the majority of the events for which the field
is $>$ 0.35 G, is shown in figure~\ref{fig:alphaplot2}(c). It is evident
that the width of the peak is larger for these events recorded under the
influence of higher transverse magnetic fields, with equal populations
for values of {\it ALPHA} between $0^\circ$ -- $15^\circ$ ($3.6 \sigma$)
and $15^\circ$ -- $30^\circ$ ($3.1 \sigma$). It should be noted that the
peak at $60^\circ$ -- $70^\circ$ is superimposed on an {\em ALPHA}-plot
with increasing frequency for {\em ALPHA} approaching $90^\circ$ --- see
Chadwick et al. (1999). The significance of this peak is therefore $\sim 2
\sigma$ (before allowing for the number of bins in the {\em
ALPHA}-plots).

\section{Conclusions}

We have demonstrated that the \v{C}erenkov images from gamma rays and
cosmic rays are broadened and rotated by the geomagnetic field. The
broadening results in an increase in the telescope threshold and a
reduction in counting rate. The rotation of the images away from the
projected direction of the magnetic field in the image plane broadens
the {\it ALPHA} distribution and decreases the sensitivity of the
telescope. 

These results suggest that the geomagnetic field can have an important
effect on the operation of atmospheric \v{C}erenkov telescopes in some
directions and that, for detected and candidate sources, the Narrabri
site is particularly susceptible. These effects should be removable using
appropriate correction techniques; unlike noise, geomagnetic effects do
not reduce the information contained in the image.

We are grateful to the UK Particle Physics and Astronomy Research
Council for support of the project and the University of Sydney for the
lease of the Narrabri site. The Mark 6 telescope was designed and
constructed with the assistance of the staff of the Physics Department,
University of Durham.

\vspace{1ex}
\begin{center}
{\Large\bf References}
\end{center}
%
Armstrong, P. et al. 1999, Experimental Astron., in press\\
Bowden, C.C.G. et al. 1992, \jpg, 18, L55\\
Chadwick, P.M. et al. 1998a, Astrop. Phys., 9, 131\\
Chadwick, P.M. et al. 1998b, ApJ, 503, 391\\
Chadwick, P.M. et al. 1999, ApJ, 513, 161\\
Hillas, A.M. 1985, Proc. 19th ICRC (La Jolla) 3, 445\\
Lang, M.J. et al. 1994, \jpg, 20, 1841\\
\end{document}